\documentclass[prb,preprintnumbers,twocolumn,amsmath,amssymb]{revtex4-1}
\usepackage{amsmath}
\usepackage{amssymb}
\usepackage{cases}
\usepackage{color}
\usepackage{txfonts}
\usepackage{graphicx}
\usepackage{wrapfig}
\usepackage{dcolumn}
\usepackage{bm}
\usepackage{wasysym}
\usepackage{textcomp}
\usepackage{braket}
\usepackage{lineno}
\usepackage{soul,xcolor}
\usepackage{natbib}

\begin{document}
\title{Strain-induced spatial and spectral isolation of quantum emitters in mono- and bi-layer WSe$_2$}
\author{S.~Kumar$^{1\ast\dag}$}
\author{A.~Kaczmarczyk$^{1\dag}$}
\author{B.~D.~Gerardot$^{1\ast}$\footnote[0]{$^{\ast}$e-mail: B.D.Gerardot@hw.ac.uk; Santosh.Kumar@hw.ac.uk}}

\affiliation{$^1$Institute of Photonics and Quantum Sciences, SUPA, Heriot-Watt University, Edinburgh EH14 4AS, UK}
\date{\today}
\begin{abstract}
Two-dimensional transition metal dichalcogenide semiconductors are intriguing hosts for quantum light sources due to their unique opto-electronic properties. Here we report that strain gradients induced by substrate patterning result in spatially and spectrally isolated quantum emitters in mono- and bi-layer WSe$_2$. By correlating localized excitons with localized strain-variations, we show that the quantum emitter emission energy can be red-tuned up to a remarkable $\sim170$ meV. We probe the fine-structure, magneto-optics, and second order coherence of a strained emitter. These results raise the prospect to strain-engineer quantum emitter properties and deterministically create arrays of quantum emitters in two-dimensional semiconductors.
\end{abstract}

\maketitle Single-photon sources are crucial for a number of emerging quantum information processing and quantum networking applications~\cite{OBrien09}. Of all the possible quantum emitters, solid-state options are leading candidates for practical applications~\cite{Santori10}. Unlike trapped ions or atoms, solid-state emitters are typically buried in a bulk, three-dimensional material which provides long-term stability. Further, semiconductors compatible with established optoelectronic processing technology enable heterostructure devices and monolithic photonic structures for enhanced functionality and performance~\cite{Lodahl15}. However, the bulk environment also provides obstacles and limitations for engineering coherent and efficient quantum light sources. First, interaction with phonons and fluctuating spins or charges within the emitter's environment can lead to dephasing~\cite{Warburton13,Gao15}. Second, a high-dielectric material presents intrinsic challenges to efficiently extract the emitted photons into a single optical mode~\cite{Lodahl15}.

An intriguing alternative to host a quantum emitter is a two-dimensional (2D) semiconductor. Like graphene, the first hexagonal Brillouin zone (BZ) of monolayer (1L) transition metal dichalcogenide (TMD) accommodates pairs of inequivalent valleys. Distinctively, the 1L TMDs offer energetic gaps at the corners of the BZ (\emph{K}-point) and strong spin-orbit coupling further splits the valence bands by hundreds of meV and the conduction bands by a few meV. This enables valley dependent optical selection rules which allow valley polarization, valley coherence, and spin-valley coupling~\cite{Xu14}. Further, strain up to $\sim11\%$ can be induced in these materials~\cite{Bertolazzi11}, raising the prospect for significant tuning of the electronic bandgap~\cite{He13experimental,Conley13,Castellanos13local,Zhu13,Desai14}. Crucially, the 2D nature eliminates the high-index environment which hinders photon extraction and is fully compatible with integrated photonics approaches~\cite{Xia14}.

Recently, localized excitons (0D-$X$) in 2D WSe$_2$ have been shown to emit non-classical light yet appear to maintain the general electronic and magneto-optical characteristics (e.g. direct band-gap, large long-range exchange interaction energy, large Coulomb screening, and large exciton g-factor) of the host 2D semiconductor~\cite{Srivastava15,He15,Koperski15,Chakraborty15,Tonndorf15}. The cause of exciton localization has been attributed to confinement at defects within the electronic band-gap of the WSe$_2$. The defects exhibit a broad-band ($\sim$60\,meV) emission feature~\cite{Tongay13} in 1L and bi-layer (2L) WSe$_2$ that appears to be energetically modulated near the edges of flakes~\cite{Koperski15,Tonndorf15} or at wrinkles~\cite{Tonndorf15}. With aggressive spectral filtering at these locations, the second-order coherence of individual emitters has been probed to demonstrate their quantum nature~\cite{Srivastava15,He15,Koperski15,Chakraborty15,Tonndorf15}. However, a general approach to obtain and potentially engineer spatially and spectrally isolated defects is essential for further development of this promising quantum photonic platform. Here we demonstrate that local strain gradients in the 2D crystal offer this capability and we take a first step towards deterministic engineering of the emitter location and optical properties.\\

\begin{figure*}[hbt] \centering
\includegraphics[width =170mm]{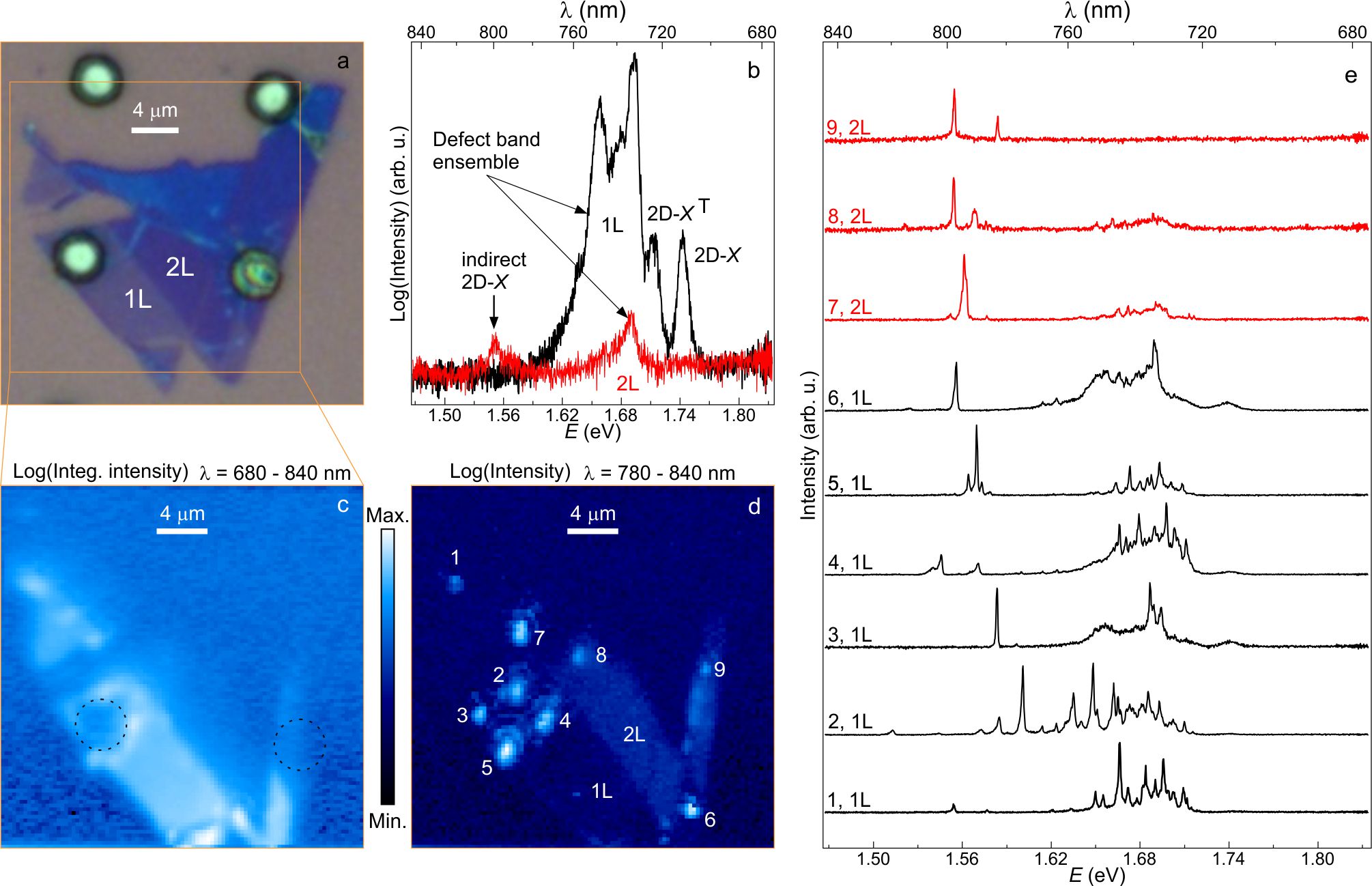}
\caption{(a)~Optical micrograph of an exfoliated flake on top of a Si substrate with etched holes (1 \textmu m deep, 4 \textmu m diameter). (b)~Typical PL emission spectra for 1L (black) and 2L (red) WSe$_2$ from the smooth and unstrained locations of the flake. The ensemble emission due to defect-bound excitonic complexes can be seen in both 1L and 2L WSe$_2$. (c-d)~Color-coded spatial maps of PL with (c) integrated intensities in the spectral range of 680-840 nm and (d) intensities in the spectral range of 780-840 nm. (e)~PL spectra corresponding to localised bright spots numbered in (d). The excitation power was 5 \textmu W. The open square in (a) marks the region of spatial maps shown in (c) and (d). The two open circles in (c) mark the locations of the etched holes. }\label{fig1}
\end{figure*}

Using an all-dry viscoelastic stamping procedure~\cite{Castellanos-Gomez14}, we transfer a mechanically exfoliated WSe$_2$ flake (from bulk, nanoScience Instruments product NS00182) onto a Si substrate pre-patterned with etched holes with 1 \textmu m depth and 4 \textmu m diameter. A confocal microscope with an objective lens with NA of 0.82, yielding a diffraction limited focus of $\sim$ 460 nm at $\lambda$ = 750 nm, was used for \textmu -PL measurements with non-resonant CW excitation at $\lambda$ = 532 nm. The microscope used dichroic mirrors at $\lambda$ = 550 nm to separate the excitation and PL signals. The sample was placed on automated nanopositioners at T = 4\,K in a closed-cycle cryostat with $B$ = 0 to 9\,T superconducting magnet.  All spectra were acquired with a 0.5 m focal length spectrometer and nitrogen-cooled charge-coupled device with a spectral resolution of $\sim$40 \textmu eV at $\lambda$ = 750\,nm for an 1800 l/mm grating. Polarization dependent PL was analysed by combining a liquid-crystal variable retarder with a quarter wave plate and fixed linear polarizer. A fiber based Hanbury-Brown and Twiss interferometer was used for second-order correlation measurements. Coincidence events from two Si avalanche photodiodes were recorded electronically on two synchronized input channels with a system timing jitter of $\sim$ 600\,ps. Efficient spectral filtering (resolution $\sim$3\,meV) was achieved using two angle-sensitive edge filters.\\

An optical micrograph of the sample is shown in Fig.~\ref{fig1}a; the grey region and four white disks represent the substrate and four etched holes, respectively, and the blue region corresponds to the WSe$_2$ flake. The two lightest contrast regions of the flake are assigned as 1L and 2L regions. A portion of the 1L flake is successfully placed onto a pre-patterned hole. Also observable in the contrast of the optical micrograph are a few wrinkles and small bumps in the flake. Non-resonant micro- photoluminescence (\textmu -PL) spectra at an excitation power of 5 \textmu W were obtained at the 1L and 2L positions of the corresponding labels in Fig.~\ref{fig1}a as shown by the  black and red curves in Fig.~\ref{fig1}b, respectively. As expected, the 1L (2L) region shows a strong (weak) PL signal. The 1L spectrum shows two high energy peaks separated by $\sim$ 31\,meV, corresponding to the quantum-well (2D) neutral exciton (2D-$X$) and a 2D charged exciton, or trion (2D-$X^{T}$)~\cite{Wang14}. The broadband emission centered at 1.68\,eV for both 1L and 2L WSe$_2$ is due to defect-bound excitons~\cite{Tongay13,Wang14}. While the 1L band-gap is direct, 2L WSe$_2$ exhibits an indirect bandgap between the conduction band minimum at $\Sigma$ and the valence band maximum at $K$ in the first BZ~\cite{Desai14}. The weak emission peak centred at 1.55\,eV in the 2L spectrum is due to this indirect 2D-$X$ transition.\\

To investigate the spatial dependence of the PL spectrum on the flake, we scanned the entire region outlined by the yellow box in Fig.~\ref{fig1}a. Figure~\ref{fig1}c shows the spatial map of the integrated intensities of the emission for 680 $< \lambda <$ 840 nm, where the 1L region of the flake has intense PL and the remaining areas are dark. A one-to-one correspondence between Fig.~\ref{fig1}a and c can be seen. The two dotted circles in Fig.~\ref{fig1}c mark the regions of holes underneath the flake. Note that the spatial map has been taken for a fixed focal plane, therefore the regions of the flake over the etched hole are significantly less bright due to alteration of the focal plane caused by buckling of the flake.  When we generate a spatial map for 780 $< \lambda <$ 840 nm, as shown in Fig.~\ref{fig1}d, then the indirect 2D-$X$ emission from the 2L region of the flake is distinctively seen. Crucially, we also observe a few  localized bright spots on the 1L flake around the edge of the etched hole and a few more isolated bright spots in both the 1L and 2L regions of the flake. These spots are numbered from 1-6 for 1L and 7-9 for 2L, and their corresponding spectra are shown in Fig.~\ref{fig1}e. The spectra from these bright spots are strikingly different from the typical emission spectra of 1L and 2L WSe$_2$ (see Fig.~\ref{fig1}b). They show a comb-like emission in the region of the defect-band and very sharp emission lines at much lower energies. While the comb-like emission has recently been reported, observation of such isolated, highly-detuned peaks is novel.\\

\begin{figure*}[hbt] \centering
\includegraphics[width =168.6mm]{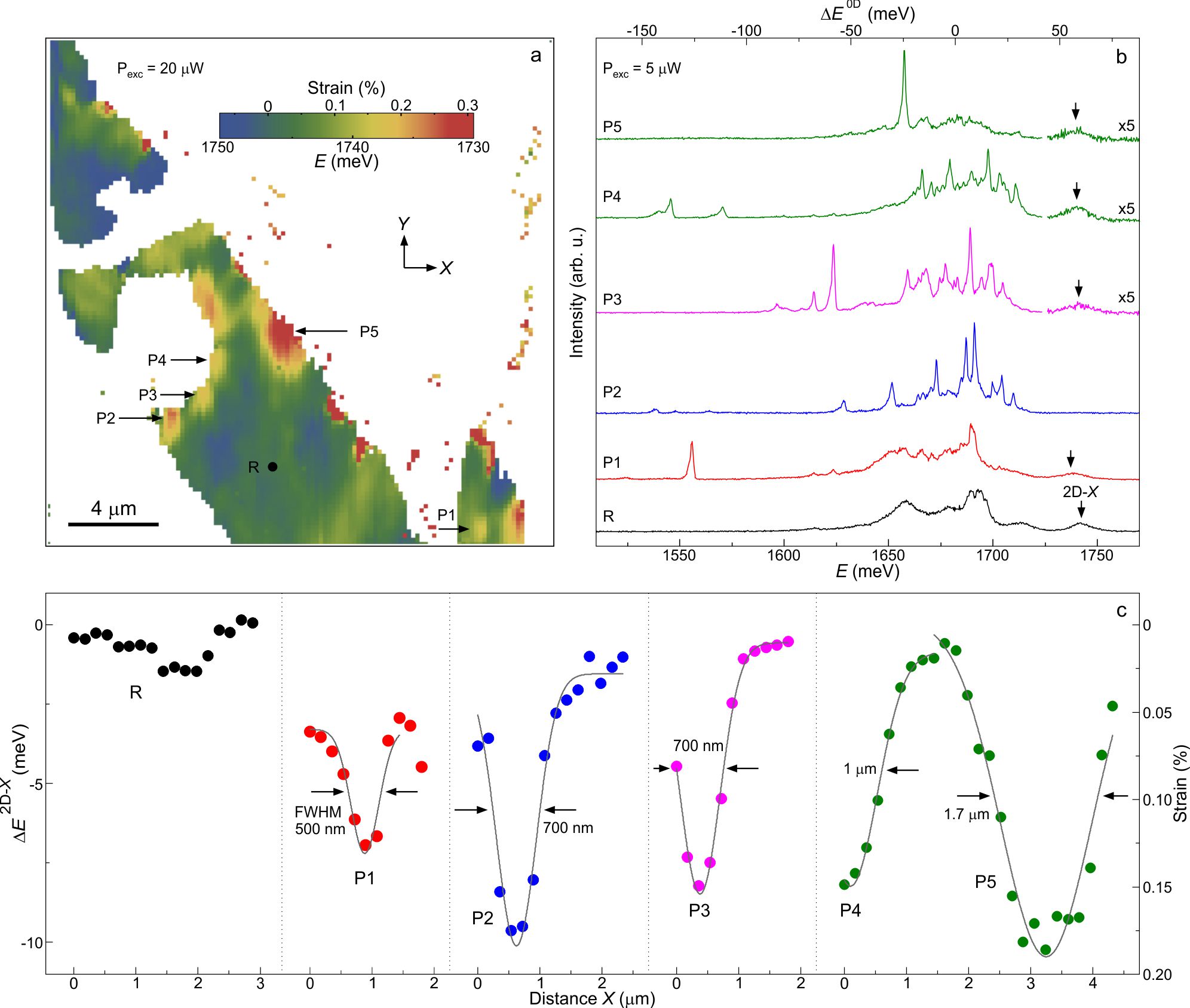}
\caption{(a)~Color-coded spatial map of the 2D-$X$ peak energy for 1L WSe$_2$ which includes a few localized energy minima. The energetic shift of the peak ($\Delta E^\text{2D-X}$) with respect to reference point R was used to estimate the strain. The excitation power was 20 \textmu W. (b-c)~(b) PL spectra and (c) horizontal cross-sectional $\Delta E^\text{2D-X}$/strain corresponding to reference point R and energy pockets P1 to P5 which correlates the effect of local strain gradients. The excitation power for (b) was 5 \textmu W.} \label{fig2}
\end{figure*}

To better understand the origin of the isolated highly red-shifted emitters, we estimate the local strain in the 1L WSe$_2$ by analyzing the peak energy of the 2D-$X$ emission at an excitation power of 20 \textmu W. The 2D-$X$ peak energy map for the 1L region of the flake is shown in Fig.~\ref{fig2}a. We choose a reference point R on 1L WSe$_2$ and estimate strain relative to this point. We use a calibration of -55 meV shift in the 2D-$X$ peak energy for 1$\%$ uniaxial tensile strain (see Fig.~S5(b) of Ref.~\cite{Desai14}). Note that strain estimation for the 2L is not possible due to a very weak indirect 2D-$X$ emission. A strain variation of $\sim-0.05\%$ (compressive) to 0.3$\%$ (tensile) across the 1L region of the flake is observed. The large portion of the 1L region of the flake which looks flat optically (see Fig.~\ref{fig1}a) shows a small strain variation ($<\pm0.05\%$). More importantly, a few highly localized energy minima are observed. We label five of these positions as P1 to P5. The region with the largest 2D-$X$ red-shift, P5, is caused by a wrinkle propagating from the 2L region as seen in the optical micrograph (Fig.~\ref{fig1}(a)). P1 corresponds to a small bump which can be seen in the optical micrograph. Pockets P2 - P4 are caused by bending of the flake around the edge of the etched hole. To characterize the spatial localization, we plot in Fig.~\ref{fig2}c the change in 2D-$X$ peak energy ($\Delta E^\text{2D-X}$) for a horizontal line-cut through the center of the five highly strained regions. Notably, pockets P1 - P3 show linewidths (full-width-half-maximum, FWHM) approaching the diffraction limit of the confocal microscope, suggesting that actual strain variations are even more localized ($<$ 500\,nm FWHM) than can be measured and are averaged out by the finite microscope resolution. Nevertheless, $\Delta E^\text{2D-X}$ (estimated strain variation) as large as $\sim$-10 meV ($\sim0.2\%$) is observed in the localized pockets.\\

The \textmu -PL spectra corresponding to the local minima P1 to P5 and R are shown in Fig.~\ref{fig2}b. The spectra at the locations show highly red-shifted sharp emission lines together with the comb-like emission of the ensemble band. In the extreme cases (see spectrum 2, 1L in Fig.~\ref{fig1}e at location 2 in Fig.~\ref{fig1}d), a red-tuning of $\sim$ 170 meV is observed relative to the mean energy of the ensemble defect band in unstrained regions. Further strain-gradient correlations are found for all emitters in Fig. 1d-e: four localized emitters occur at the edge of the patterned hole, the remaining occur at locations of unintentional strain. Conversely, the spectrum at R and elsewhere on the flat 1L region shows only a broad band 0D-$X$ ensemble (not shown). This direct correlation of localized strains and spatial and spectral emitter isolation leads to a robust conclusion that local strain variations disperse the energies of the individual defects in the 0D-$X$ ensemble, resulting in comb-like emission within the nominal defect band and much larger red-tuning of a few defects at highly strained positions. Though we estimate a maximum strain of only $\sim$0.3$\%$, we believe that larger strains exist at more highly localized spatial positions than we can measure because of the finite microscope resolution.\\

\begin{figure}[hbt] \centering
\includegraphics[width =85.82mm]{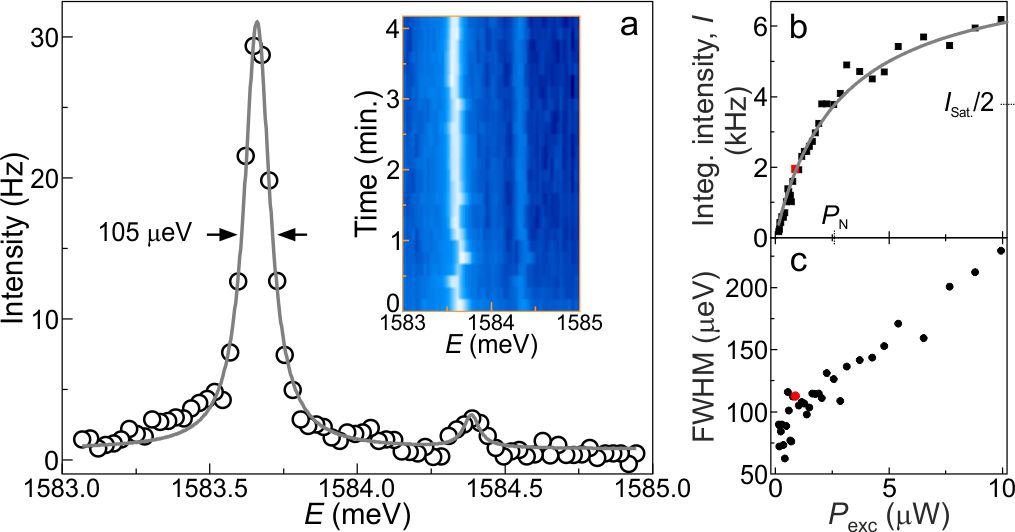}
\caption{(a)~High resolution PL spectrum of 0D-$X$ from location 3 of Fig. 1d. The experimental data is fit by two Lorentzian curves (grey line). The two peaks of the doublet are fine-structure-split 0D-$X$ lines and the unequal peak intensities are due to anisotropic local strain. Inset: Spectral fluctuations of the doublet revealed using a color-coded time trace of the 0D-$X$ lines. The time bin is 10 seconds. (b-c)~The excitation power dependences of (b) integrated intensity and (c) linewidth of the low-energy 0D-$X$ line. The grey line is the saturation curve fit. The red data points in (b) and (c) mark the representative positions for the spectrum shown in (a).} \label{fig3}
\end{figure}

The \textmu -PL emission spectrum at location 3 on 1L WSe$_2$ flake (see Fig.~\ref{fig1}e) shows a sharp and spectrally isolated emitter at $\lambda$ = 782.72\,nm. High-resolution spectroscopy reveals a doublet split by $\Delta$ = 726\,\textmu eV with unequal intensities, as shown in Fig.~\ref{fig3}a. Each component of the doublet shows saturation and an excitation power ($P_\text{exc}$) dependent linewidth (see Fig.~\ref{fig3}b and c, respectively). The solid line in Fig.~\ref{fig3}b is a saturation curve fit for a two-level system using the relation $I=I_\text{sat}\left(P_\text{exc}/\left(P_\text{exc}+P_\text{N}\right)\right)$, where $P_\text{N}$ = 2.78\,\textmu W is the normalization excitation power at which integrated intensity of the emission peaks becomes half of the saturation integrated intensity ($I_\text{sat}$). It shows a linear power dependence at low $P_\text{exc}$ and a clear saturation behavior at high $P_\text{exc}$. At the lowest $P_\text{exc}$, the minimum linewidths measured are $\sim$ 60\,\textmu eV (FWHM). With increasing $P_\text{exc}$, increasing inhomogeneous broadening is observed. The inhomogenous broadening is likely caused by fluctuating charges in the environment of the quantum emitter at a time scale faster than the experimental acquisition time. Also, modest spectral wandering is observed at longer time scales, as shown in the inset of Fig.~\ref{fig3}a. A naive expectation for a quantum emitter exposed at the surface is severe non-radiative recombination, inhomogeneous broadening, and photobleaching caused by nearby surface states. Surprisingly, the 0D-$X$ states in WSe$_2$, perhaps aided by the strong exciton binding energy, do not exhibit such deleterious features. Further, we find the amount of inhomogeneous broadening is directly linked to the non-resonant excitation power (Figure~\ref{fig3}(c)), suggesting that quasi or strictly resonant excitation could lead to minimal electric field fluctuations and nearly transform limited linewidths for 0D-$X$ states.\\

\begin{figure}[hbt] \centering
\includegraphics[width =86.04mm]{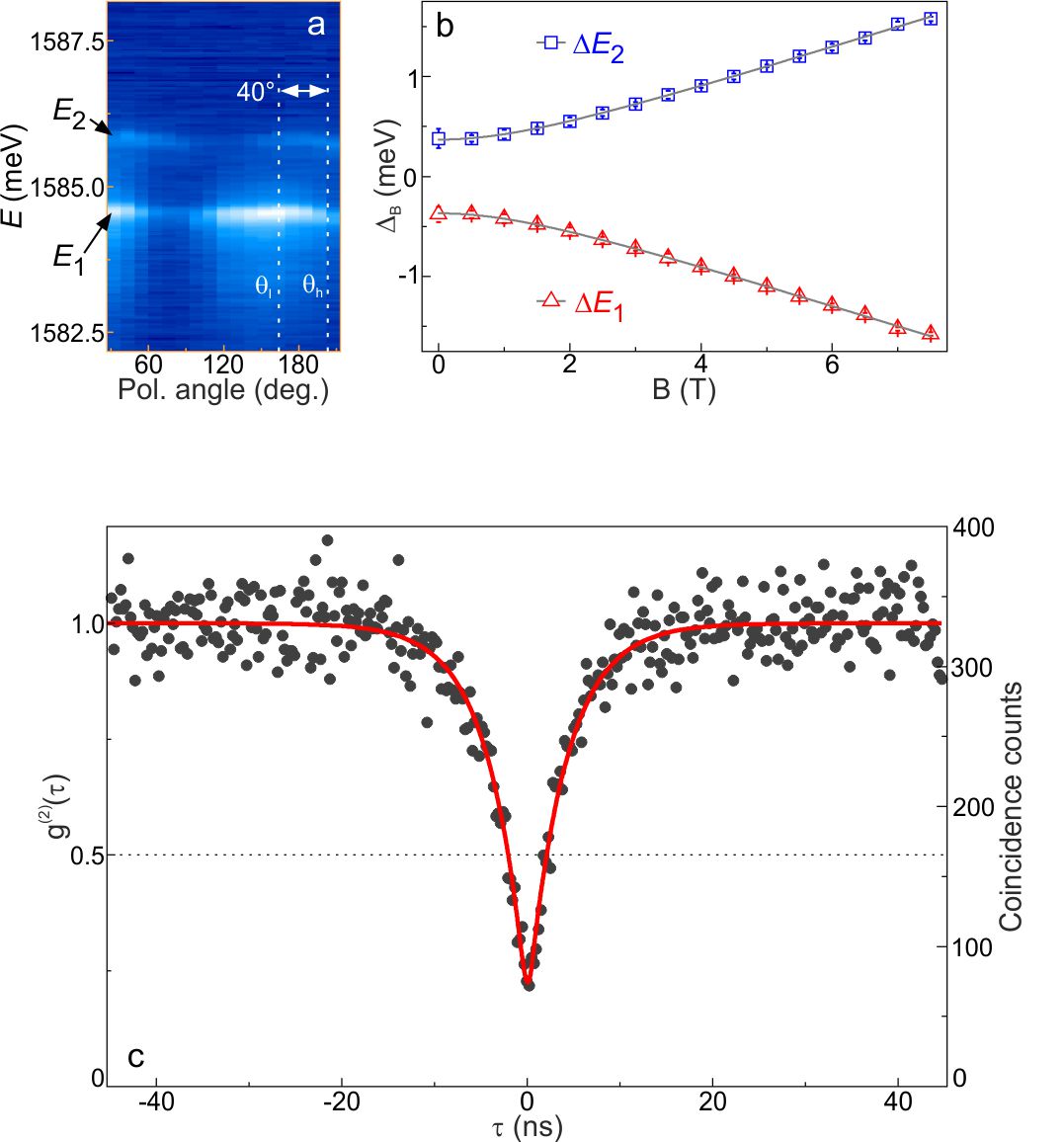}
\caption{(a)~Color-coded PL intensity as a function of polarization of the 0D-$X$ lines at $B$ = 2.5 T in Faraday geometry. The two dotted vertical lines show polarization direction of low-energy ($\theta_\text{l}$) and high-energy ($\theta_\text{h}$) 0D-$X$ lines. A 40\textdegree angle between these directions also reflects the presence of anisotropic local strain. (b)~Magnetic field dependence of the fine-structure split doublet.  (c)~Normalized second-order correlation function $g^{(2)}(\tau)$ of the 0D-$X$ lines under non-resonant cw excitation at $P=0.5P_\text{N}$. The fit (solid line) shows $g^{(2)}(0)$ = 0.17 and a lifetime of 4.14 ns.} \label{fig4}
\end{figure}

To investigate the nature of the doublet, we perform polarization-resolved and magneto-optical spectroscopy. Figure~\ref{fig4}a shows a polarization resolved intensity map of the doublet, revealing that the lines have unequal maximum intensities, are linearly polarized, and the angle between their polarization directions is 40\textdegree . We attribute the doublet to be the fine-structure of a neutral 0D-$X$. The symmetry breaking of the confinement potential leads to mixing of the two neutral excitonic states due to the electron-hole spin-exchange interaction, and therefore, the emission splits into two lines. The energetic separation between these two lines is called the fine-structure splitting ($\Delta$)~\cite{Gammon96,Bayer02}. Here we observe $\Delta=726$\,\textmu eV. Similar values of $\Delta$ were reported recently~\cite{Srivastava15,He15,Koperski15} and the relatively large magnitude is likely due to the large long-range exchange interaction energy. Usually, two lines of the fine-structure split doublet have equal maximum intensities and are linearly polarized along orthogonal directions. The unequal intensities of the doublet lines and the non-orthogonality between their polarization directions of the emitter at location 3 can be ascribed to the presence of anisotropic strain~\cite{Kumar14}. We have also performed polarization dependent PL for other emitters at locations with much less strain and find that the lines of the doublets show nearly equal maximum intensities and their polarization directions are nearly orthogonal to each other (not shown).\\

The fine-structure split doublet's behavior in an external magnetic field ($B$) is illustrated in Fig.~\ref{fig4}b. In the Faraday geometry (direction of magnetic field perpendicular to the flake plane), $\Delta$ is modified according to $\Delta_\text{B}=\sqrt{\Delta_\text{0}^{2}+\left(\mu_\text{B}g_{\text{0D}-X}B\right)^{2}}$, where $\mu_\text{B}$ is the Bohr magneton and $g_{\text{0D}-X}$ is exciton g-factor. The clean fit of the data to $\Delta_\text{B}$ affirm that both lines of the doublet originate from a single neutral 0D-$X$; the fit reveals $g_{\text{0D}-X}$\,=\,7.16\,$\pm$\,0.02. The large $g_{\text{0D}-X}$ is not understood, but is in agreement with recent reports~\cite{Srivastava15,He15,Koperski15,Chakraborty15}. The large $g_{\text{0D}-X}$ inspires  further investigations into the spin-valley degree of freedom in the TMD quantum emitters. Additionally, $B$ also changes the central emission energy ($E_\text{Avg.}$) of the doublet due to the diamagnetic shift~\cite{Walck98}, which is given by $E_\text{Avg.}\left(B\right)=E_\text{Avg.}\left(0\right)+\gamma\left(B\right)^{2}$, where $\gamma$ is the diamagnetic coefficient. For this 0D-$X$, we measure $\gamma$ = -3.9$\pm$1.0\,\textmu eV/T$^{2}$ (not shown). For a 3D confined system, $\gamma$ quantifies the combined contributions of confinement and Coulomb interaction upon application of the magnetic field~\cite{Walck98}. The very small $\gamma$ observed here demonstrates very strong confinement of 0D-$X$.\\

Finally, we observe strong anti-bunched photon emission from this red-shifted emitter. Figure~\ref{fig4}c presents the second-order correlation function $g^\text{(2)}(\tau)$ under non-resonant cw excitation ($P = 0.5 P_N$). The closed circles are measured data and the solid line is the fit using the relation $g^\text{(2)}(\tau)$\,=\,$1-\rho^{2}e^{-|\tau|/T_{1}}$, where $T_{1}$ is the lifetime and \textit{SBR}\,=\,$\rho/(1-\rho)$ is the signal-to-background ratio (\text{SBR}). We obtain $g^\text{(2)}(\tau)$\,=\,0.17\,$\pm$0.02 from the fit, unambiguously proving quantum emission. The fit yields $T_{1}$\,=\,4.14\,$\pm$0.15 ns and the estimated \text{SBR} here is $\sim$10.\\

In summary, we have achieved resolution-limited spatial localization of 0D-$X$ with extremely low densities of $\sim$ 1 $\mu$m$^{-2}$ within a 50 nm emission bandwidth (760 $< \lambda <$ 810 nm) and tuning of the 0D-$X$ emission energy over huge range (up to $\approx$ 170 meV). We directly correlate the 0D-$X$ tuning and spatial and spectral isolation to microscopic pockets of large strain variation. These results demonstrate that strain engineering is a viable approach to obtain spatially and spectrally isolated quantum emitters in 2-D semiconductors. The passive exciton emission tuning observed with the 2D quantum emitters is significantly larger than the state-of-the-art for quantum emitters in bulk semiconductors~\cite{Kumar14,Ding10,Trotta12,Kuklewicz12}. However, there is significantly more potential for strain engineering 0D-$X$ states. First of all, massive strain gradients are possible in 2D materials~\cite{Bertolazzi11} and a number of approaches for both static or \textit{in-situ} tunable uniaxial and biaxial strain engineering can be considered~\cite{Roldan15}.  Secondly, the affect of strain on the electronic and optical properties of 0D-$X$ states beyond their emission energy has yet to be investigated. For instance, \textit{in-situ} strain tuning can enable engineering of the carrier confinement potential~\cite{Kuklewicz12}, the permanent dipole moment, the fine-structure splitting~\cite{Kumar14,Ding10,Trotta12}, and the spin properties~\cite{Huo14natp} of localised exciton states. Notably, such quantum emitters could underpin a hybrid semiconductor-atomic interface~\cite{Akopian11np,kumar11}. Finally, rather than suspending the flake over a large hole, by placing the flake over a nanostructured surface such as a periodic array of steps or holes, one can envisage periodic arrays of strained quantum emitters.\\

We thank A. Castellanos-Gomez for sharing his expertise on the dry transfer technique, R.  Bernardo-Gavito and D. Granados for assistance assembling the microscope, J.M. Zajac for assistance with sample imaging, and A. Rastelli for data analysis software. This work was supported by a Royal Society University Research Fellowship, the EPSRC (grant numbers EP/I023186/1 EP/K015338/1, and EP/L015110/1) and an ERC Starting Grant (number 307392).\\

$^\dag$SK and AK contributed equally to this work. Corresponding Author $^\ast$E-mail: Santosh.Kumar@hw.ac.uk; B.D.Gerardot@hw.ac.uk.\\

\end{document}